\documentclass{aa}  
\usepackage{hyperref}
  
\usepackage{cleveref}
\usepackage{dblfloatfix}
\usepackage{amsmath}
\usepackage{graphicx}
\usepackage{txfonts}
\usepackage{lipsum}
\usepackage{subcaption}         % necessary for continued figures, example in section 3
\usepackage{textgreek}
\usepackage{balance}                                % and appendix
\usepackage{lscape}             % to rotate a single page table, example in appendix.
\hypersetup{
  colorlinks   = true,
  linkcolor    = blue,   
  citecolor    = blue,   
  urlcolor     = blue
}                              % For landscape tables, see the longtable examples.
\usepackage{placeins}           % useful with \FloatBarrier, to keep 
\usepackage{tikz}
\usetikzlibrary{positioning}
\usetikzlibrary{arrows}
\DeclareMathOperator\erfc{erfc}
\usepackage{orcidlink}
\usepackage{hyperref}
\begin{document}

%%%%%%%%%%%%%%%%%%%%%%%%%%%%%%%%%%%%%%%%
% if you use custom commands in your title,
% ensure to check your title when submitting!
%%%%%%%%%%%%%%%%%%%%%%%%%%%%%%%%%%%%%%%%
   \title{Detection of cosmic strings by gravitational wave lensing. Predictions for Einstein Telescope}

  % \subtitle{Subtitle}

%%%%%%%%%%%%%%%%%%%%%%%%%%%%%%%%%%%%%%%%
% Please separate each author with the \and command
%
% Please do not include ORCIDs next to author names.
% Only ORCIDs authenticated by individual authors in EDPS
% editorial system will be taken into account.
% ORCIDs included here will be removed.
%%%%%%%%%%%%%%%%%%%%%%%%%%%%%%%%%%%%%%%%

\author{
J.~Szyndler\inst{1,2}
\and
T.~Bulik\inst{2}
}

\institute{
Faculty of Physcis, University of Warsaw,
Pasteura 5, 02-093  Warsaw, Poland
\and
Astronomical Observatory, University of Warsaw,
Al. Ujazdowskie 4, 00-478 Warsaw, Poland
\email{j.szyndler@student.uw.edu.pl}
\email{tb@astrouw.edu.pl}
}
   %\date{Received July 07, 2026}

% \abstract{}{}{}{}{}
% 5 {} token are mandatory
 
  \abstract
  % context heading (optional)
  % {} leave it empty if necessary  
   {Cosmic strings are not yet confirmed, topological defects formed in the early Universe.  They can bend light or gravitational waves, which causes an effect similar to the gravitational lensing. }
  % aims heading (mandatory)
   {Our goal is to check whether cosmic string could be detected as a lenses of gravitational waves by the Einstein Telescope (ET).} 
  % methods heading (mandatory)
   {To do that the apparatus of wave optics had been applied. Firstly we explored the amplification factor strength and behaviour. Next the wave effects in SNRs and waveforms was examined for different inclinations of the source. Lastly  we estimated the string tensions based on eight mergers from the \textsc{StarTrack} simulation, using the Bayesian Inference methods.}
  % results heading (mandatory)
   {The wave effects were easily to see in waveforms, SNR and characteristic strains. Also most of the events could be detected, based on their Signal to Noise Ratio values. Almost whole characteristic strain lies in the range of the ET. When it comes to mock observations, we had got estimated vale of a logarithm of the CS tension equal to $\bar{\log{G\mu}}=-9.80^{+0.49}_{-0.56} $, which were consistent with injected value equal to $-10$.  }
  % conclusions heading (optional), leave it empty if necessary
   {At the end we conclude that CSs could be detected by the ET.}


   \maketitle
    \nolinenumbers

%%%%%%%%%%%%%%%%%%%%%%%%%%%%%%%%%%%%%%%%%%%%%%%%%%%%%%%%%%%%%%
\section{Introduction}\label{sec:intr}

Cosmic strings (CS) are possible one-dimensional topological defects, predicted by some gauge theories \citep{Kibble:1976sj,1985PhR...121..263V,1993ppc..book.....P}. Formed in the early Universe, usually during inflation, they are products of spontaneous symmetry breaking mechanisms \citep{1984PhRvD..30.2036V}. Cosmic strings can occur in two forms: loops or very long (super)strings \citep{1985PhLB..153..243W,2010RSPSA.466..623C}. This work focuses on the latter. One of the fundamental parameters of cosmic strings is the dimensionless tension ($G\mu$), which can vary over the range ${G\mu} < 10^{-3}$ \citep{Witten:1984eb}. Current studies constrain it to $\sim10^{-7}$ from Cosmic Microwave Background (CMB) measurements by \textit{Planck} \citep{2014A&A...571A..25P} and to ${G\mu} \sim10^{-11}$ from gravitational wave observations \citep{10.1093/mnras/stv1538,2021PhRvL.126x1102A}.

In the past, there were various attempts to find signatures of cosmic strings. Phenomena such as gravitational wave (GW) signals \citep{2021PhRvD.103j3512B, 2021PhRvL.126x1102A, 2023PhRvD.108j3511E, 2024PhRvD.109b2006M}, CMB anisotropies \citep{2017JCAP...06..004H}, repetition of FRBs \citep{2022PhRvD.106j3033X}, femtolensing of gamma-ray bursts \citep{2013PTEP.2013a3E01Y}, and optical gravitational-lensing signatures \citep{2003MNRAS.343..353S,2008MNRAS.385.1959G, 2010MNRAS.406.2452M, 2011PhRvD..83l2004C,2014PhRvD..89l4003B} were considered. One such case was the CSL-1 lens \citep{2003MNRAS.343..353S, 2005astro.ph.11085F, 2007MNRAS.376.1731S}, a galaxy image pair that turned out to be a galaxy merger \citep{2006PhRvD..73h7302A}. Cosmic strings were also considered as an origin of large-scale structure \citep{1991PhST...36..114B}. However, constraints from the \textit{Planck} satellite dismissed this theory \citep{2014A&A...571A..25P}. To date, the existence of cosmic strings remains unresolved.

In \cite{2008MNRAS.384..161K}, it was shown that the maximum possible number of cosmic strings in the intergalactic medium that could be detected by microlensing of quasars is $N_{\max}\sim3\times10^5$, constrained by current observations, with a detection rate of $\Gamma_{\max}\sim1.5\times 10^{-6}\,\text{year}^{-1}$. Also, according to \cite{2020MNRAS.491..596C}, typical survey strategies are sensitive to detecting CS via lensing of Type~Ia supernovae or stars in M31 (the Andromeda galaxy). However, these experiments should have a very long cadence.

Lensing phenomena are not limited to light; gravitational waves (GW) can also be lensed \citep{2003ApJ...595.1039T, 2023Univ....9..200G}. Other "optical" effects, such as interference \citep{2025SSRv..221...29L} or diffraction \citep{tkk2-x9st}, could be observed. In this case, we deal with wave optics beyond the geometrical-optics regime if the wavelength is much larger than the Schwarzschild radius of the lens \citep{2018PhRvD..98j4029D}. Searches for lensing effects in GW data were conducted \citep{2025arXiv251216347T}. There are reasons to suppose that the first event of gravitational waves lensed by a point mass was GW231123 \citep{2025ApJ...993L..25A, 2025arXiv251217631G}. Tools to detect this phenomenon have been developed and are still being improved \citep{2025PhRvD.111j3539V, 2026PhRvD.113b4041C}. Gravitational wave lensing could also be a useful tool for dark-matter searches \citep{2019PhRvL.122d1103J, 2020MNRAS.495.2002L, 2021MNRAS.502L..16C, 2022MNRAS.509.1358U, 2024PhRvD.109l4020C, 2025PhRvD.111b4068B, 2025JCAP...07..025S, 2025arXiv251014953V} or testing general relativity \citep{2020PhRvD.102l4048E, 2023PhRvD.108b4052G, 2024PhRvD.109l4014H, 2025PhRvD.112b4073S}.

The lensing of GWs by cosmic strings has been considered \citep{PhysRevD.73.024026, 2020JCAP...07..068J, 2016PhLA..380.2897F, 2022JCAP...07..022B, 2025arXiv251020442B}. These studies show that the presence of cosmic strings should manifest through observable distortions, such as double images or characteristic patterns in a single image. In \citep{2025arXiv251020442B}, the possibility of distinguishing a CS lensing signal from a point-mass signal was investigated. The authors found that this phenomenon could be detected even with the LIGO--Virgo--KAGRA configuration. This opens a new frontier for future detectors to conduct novel searches for cosmic strings.

In this work, we use the Einstein Telescope as a model detector \citep{2010CQGra..27s4002P}, a European project for a third-generation gravitational wave (GW) observatory. Its objectives are broad and cover many science cases, such as cosmology and compact-object population studies \citep{2023JCAP...07..068B, 2025arXiv250312263A}. Here, we study whether it could be used as a tool to search for cosmic strings via a previously proposed method. The paper is organized as follows. In Section~\ref{sec:formalsim}, we introduce the lens model and wave effects. Next, in Section~\ref{sec:amp}, we describe these effects in terms of the amplification factor. In Section~\ref{sec:detect}, we discuss detection based on individual lensing events. Finally, in Section~\ref{sec:ET_sim}, we propose a method to estimate the cosmic-string tension using multiple lensing events. The discussion and summary are given in Section~\ref{sec:disc_summ}.

\section{Gravitational Wave lensing by cosmic strings}\label{sec:formalsim}

If the wavelength is much larger than the Schwarzschild radius of the lens, diffraction becomes important \citep{2018PhRvD..98j4029D} and the magnification becomes very small \citep{1986ApJ...307...30D}. Interference must also be taken into account, since gravitational waves are coherent \citep{1986PhRvD..34.1708D}. Thus, one often cannot solve the GW lensing problem in the geometric-optics regime. These wave effects force us to use a different methodology than in the electromagnetic case; we adopt the \textit{wave optics} formalism \citep{1992grle.book.....S}.

One of the fundamental quantities in wave optics is the amplification factor \citep{1999PThPS.133..137N,2025PhRvD.111j3539V}:
\begin{equation}
    \label{equ:F_def}
    F(f) \equiv \frac{\tilde{h}(f)}{\tilde{h}_0(f)}\,.
\end{equation}
Here, $\tilde{h}_0(f)$ and $\tilde{h}(f)$ are the characteristic strains in the frequency domain for the unlensed and lensed gravitational wave signals, respectively.

As shown in \citep{2019GReGr..51..160H}, to discuss the lensing problem we need the background metric, which can be written as \citep{1985ApJ...288..422G,1985PhR...121..263V}
    \begin{equation}
    \label{equ:string_metric}
        ds = -dt^2+dz^2+dr^2+(1-4G\mu)^2r^2d\varphi^2
    \end{equation}
As one can see, it is possible to locally recover the Minkowski metric with the change of variables $\varphi' = (1-4G\mu)\varphi$. This indicates a conical geometry (a ``folding'' of spacetime), which can produce a double signal from the same source. From \citep{1984ApJ...282L..51V,PatrickPeter_1994,2007MNRAS.376.1731S,PhysRevD.37.3438}, we know that in our case the separation between these two images is constant and equal to
    \begin{equation}
        \label{equ:separation_angle}
        \Delta = 8\pi G \mu
    \end{equation} 
 Based on this information, the characteristic angle (Einstein angle) $\theta_\ast$, which tells us about strength of the event, can be derived \citep{2022PhRvD.106j3033X}.
    \begin{equation}
        \label{equ:einstein_angle}
        \theta_\ast = 4\pi G \mu \frac{D_{LS}}{D_S}
    \end{equation}
\begin{figure}
    \centering
    \includegraphics[width=0.95\linewidth]{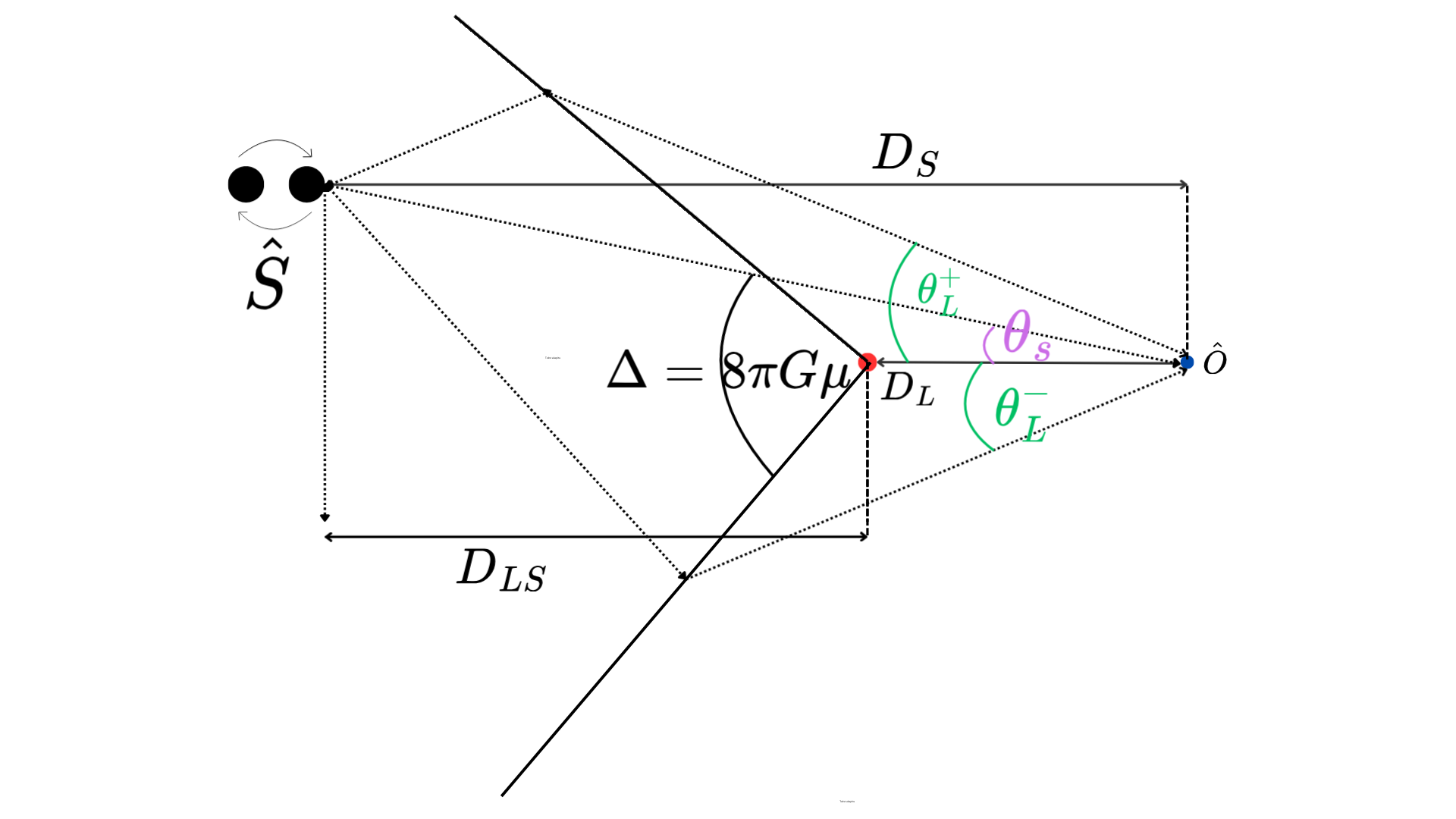}
    \caption{Gravitational wave lensing by a cosmic string. The variables are as follows: $\theta_L^\pm$ is the angular position of the lensed images; $\theta_S$ is the angular position of the source; $D_S$ and $D_L$ are the angular-diameter distances to the source and lens, respectively; and $D_{LS}=D_S-D_L\frac{1+z_L}{1+z_S}$ is the distance between the source and the lens.}
    \label{fig:scheme}
\end{figure}
 For making easier the calculation we define the effective distances
    \begin{align}
    D_\mathrm{eff} &= \frac{D_LD_{LS}}{(1+z_L)D_S} \label{equ:d_eff}\\
    \chi_{eff} &= \chi_L(1-\frac{\chi_L}{\chi_S}) \label{equ:chi_eff}    
    \end{align}
where $D_L$, $D_S$, and $D_{LS}=D_S-D_L\frac{1+z_L}{1+z_S}$ are the angular-diameter distances defined in Fig.~\ref{fig:scheme}, and $\chi_L$ and $\chi_S$ are the comoving distances to the lens and source, respectively.

Next, we define dimensionless quantities following \citep{1999PThPS.133..137N, 2013PTEP.2013a3E01Y,2025PhRvD.111j3539V}:

\begin{align}
    \varphi &= \theta_S-\theta_\ast, \\
    y &= \frac{\varphi}{\theta_\ast} \label{equ:y} \\
    w &= 2\pi\frac{(D_L\theta_\ast)^2}{D_\mathrm{eff}}f \label{equ:w}
\end{align}
Here, $\theta_S$ is the angular position of the source, $\theta_\ast$ is the characteristic angle defined in Eq.~\ref{equ:einstein_angle}, $D_\mathrm{eff}$ is the effective distance defined in Eq.~\ref{equ:d_eff}, $y$ is the dimensionless source position, and $w$ is the dimensionless frequency.

Now we focus on the amplification factor derived in \citep{2003PhRvD..68d1302Y,PhysRevD.73.024026}. We use the simplified form from \citep{2013PTEP.2013a3E01Y,2020JCAP...07..068J}:
\begin{equation}
\label{equ:F_approx}
\begin{split}
F(w,y)=\exp(\frac{w}{2i}(1+2y))[1-\frac{1}{2}\erfc{(\sqrt{\frac{w}{2i}}(1+y))}]+ \\
\exp(\frac{w}{2i}(1-2y))[1-\frac{1}{2}\erfc{(\sqrt{\frac{w}{2i}}(1-y))}]
\end{split}
\end{equation}
where $\erfc(x)=\frac{2}{\sqrt{\pi}}\int_x^\infty e^{-t^2}\,dt$.

To ensure that the arrival time of the first wavefront is zero, we follow \citep{2020JCAP...07..068J} and introduce an extra phase shift,
\begin{align}
    \phi_m = \frac{w}{2}-wy\,. \label{equ:extra_phase}
\end{align}
With this quantity, we can apply a correction to Eq.~\ref{equ:F_def}:
\begin{equation}
\label{equ:F_phased}
    \tilde{h}(w,y)=\tilde{h}_0(f)F(w,y)\exp(i\phi_m)\
\end{equation}
Here, $F(w,y)$ is the amplification factor from Eq.~\ref{equ:F_approx}, $\phi_m$ is the extra phase shift from Eq.~\ref{equ:extra_phase}, and $\tilde{h}_0$ and $\tilde{h}$ are the unlensed and lensed waveforms in the frequency domain.

References \cite{2020JCAP...07..068J,2022JCAP...07..022B} showed that, during GW lensing, characteristic fringes occur at the following frequencies \citep{2020JCAP...07..068J}:
\begin{equation}
\label{equ:f_fringe}
    f_\text{fringe} = \frac{D_{eff}}{2y(D_L\theta_\ast)^2}
 \end{equation}
%With all the derived formulas of  necessary quantities, one can move further to calculate waveforms and signal to noise ratio of lensed gravitational waves from binary black holes mergers. 

We can also obtain the time delay between the images (and its inverse), following \citep{2025arXiv251020442B}:
\begin{equation}
    \label{equ:time_delay}
    \Delta t =32y\pi^2\chi_{\mathrm{eff}}(G\mu)^2=32y\pi^2\mu_{\mathrm{eff}}^2\,.
\end{equation}
Here, $\mu_{\mathrm{eff}}$ is an effective tension, defined as
\begin{equation}
    \label{equ:mu_eff}
    \mu_{\mathrm{eff}}=G\mu\sqrt{\chi_{\mathrm{eff}}}\
\end{equation}
Knowing the time difference between the wavefronts helps constrain the lens parameters.

The method described above allows us to assess whether as-yet-undiscovered cosmic strings could be detected via gravitational wave lensing.

\section{Wave effects in amplification factor }\label{sec:amp}
Because of their nature, gravitational wave lensing involves wave effects such as interference and diffraction \citep{2025SSRv..221...29L,tkk2-x9st}. These effects are visible in the waveforms (in both the time and frequency domains), in the amplification, and in the signal-to-noise ratio (SNR) \citep{2003ApJ...595.1039T}.

\begin{figure}
    \centering
    \includegraphics[width=0.975\linewidth]{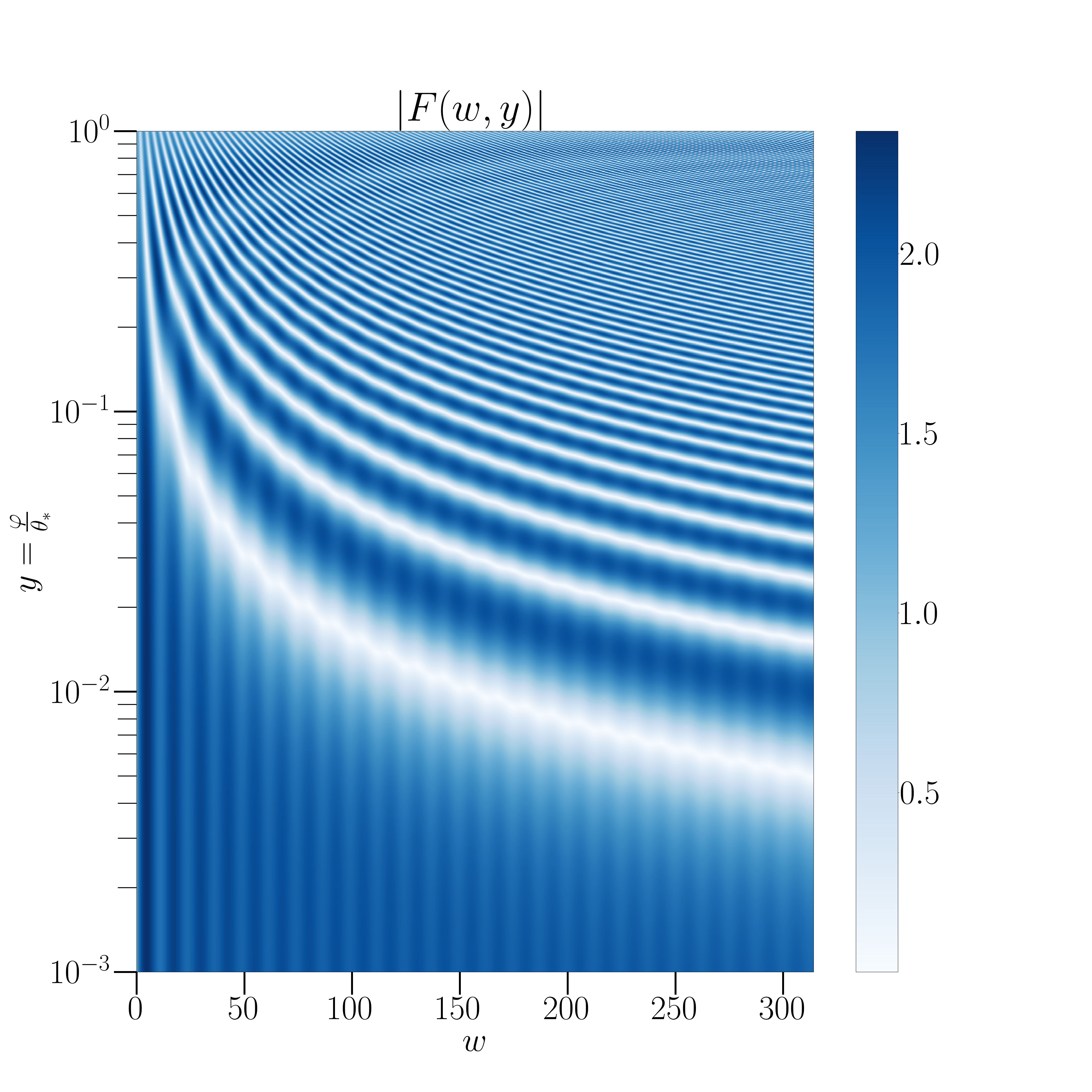}
    \caption{Amplification factor as a function of $w$ and $y$. The horizontal axis shows the dimensionless frequency $w$, and the vertical axis shows the dimensionless source position $y$. The colour bar indicates the absolute value of the amplification factor $|F(w,y)|$.}
    \label{fig:amp_2D}
\end{figure}

Let us focus on the amplification factor, which depends on the observer position and the source frequency. It is given by Eq.~\ref{equ:F_approx}. Based on this, we construct a 2D heat map of $F(w,y)$ (Fig.~\ref{fig:amp_2D}).

The highest values occur at larger $w$ and smaller $y$, indicating an interference-dominated regime. Characteristic diffraction fringes appear for $y\gtrsim 0.005$ and extend over the full range of $w$. The fringes become denser as $y$ and $w$ increase, indicating oscillatory behaviour typical of diffractive effects. We also observe regions where the amplification is strongly suppressed, characteristic of destructive interference. For $y\lesssim 0.005$, the signal is predominantly amplified.

\begin{figure*}
    \centering
    \includegraphics[width=0.97\linewidth]{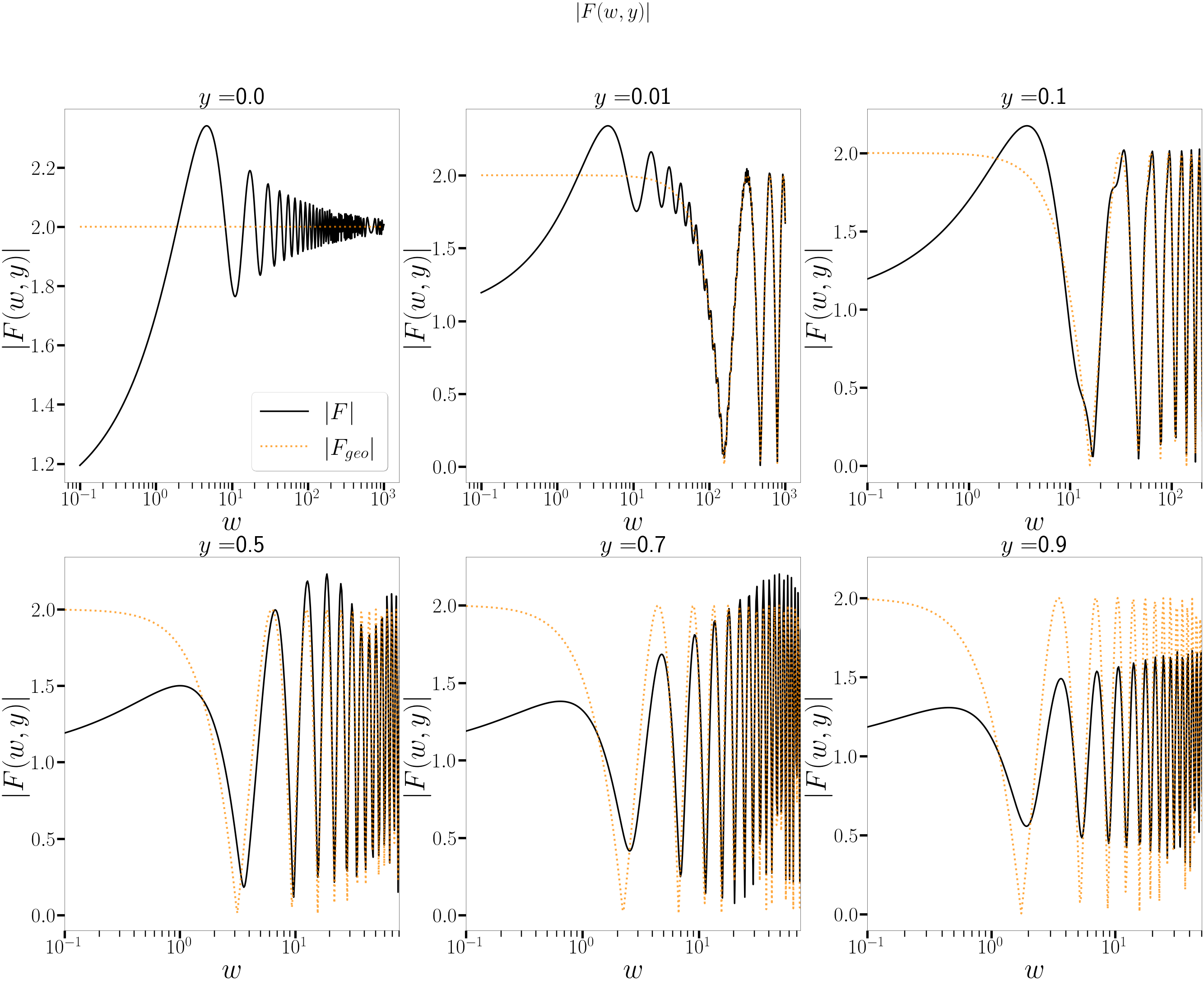}
   \caption{Absolute value of the amplification factor for $y=0,0.01,0.1,0.5,0.7,0.9$ (from upper left to lower right). The solid black curve shows $|F(w,y)|$ from Eq.~\ref{equ:F_approx}. The dashed orange curve shows the geometric-optics approximation $|F_{\mathrm{geo}}(w,y)|$ from Eq.~\ref{equ:F_geo}.}
    \label{fig:F_w}
\end{figure*}

Next, we consider standard plots of the amplification factor versus the dimensionless frequency $w$. To analyse the behaviour more precisely, we select vertical slices of Fig.~\ref{fig:amp_2D} at $y\in\{0,0.01,0.1,0.5,0.7,0.9\}$. For comparison with the geometric-optics approximation, we use the expression from \citep{2013PTEP.2013a3E01Y}: 

\begin{equation}
    \label{equ:F_geo}
    F_{geo} = \sqrt{2+2\cos(2wy)}
\end{equation} 
where $w$ is the dimensionless frequency defined in Eq.~\ref{equ:w}, and $y$ is the dimensionless source position defined in Eq.~\ref{equ:y}. The results are presented in Fig.~\ref{fig:F_w}.

We can now clearly see how the wave-optics regime differs from geometric optics. The oscillations in $|F|$ become denser as $w$ increases. For $y>0$, the diffractive behaviour is clearly visible, and $|F|$ oscillates around unity. For $y\approx 1$, the geometric-optics approximation overestimates the amplification. For $y=0$, the amplification oscillates around $|F_{\mathrm{geo}}|=2$. These results indicate that, for cosmic-string lensing, wave optics is required and the geometric-optics approximation is insufficient.

\section{Cosmic string detection}\label{sec:detect}

The main task of this work is to assess whether cosmic strings could be detected via gravitational wave lensing. To do this, we assume an infinitely long straight string with tension $G\mu\approx 10^{-10}$, motivated by current constraints \citep{2014A&A...571A..25P,10.1093/mnras/stv1538,2021PhRvL.126x1102A}. We also assume a spatially flat $\Lambda$CDM universe with parameters consistent with the \textit{Planck} results \citep{2020A&A...641A...6P}.
%\newpage
\subsection{Model of cosmic string}\label{subsec:model}

In this section, we choose representative parameters for the cosmic-string model. As discussed in Section~\ref{sec:intr}, the dimensionless tension is constrained by current observations. In \cite{2025arXiv251020442B}, the separation angle was taken to be $\Delta\theta=10^{-8}$. Since we are examining the capabilities of the Einstein Telescope, which is expected to be more sensitive than the LVK detectors, we adopt $G\mu\approx10^{-10}$. This value is still allowed by current constraints (e.g., from pulsar timing arrays; see Section~\ref{sec:intr}).

To fully explore the wave effects of examined cosmic strings we assume that the lens lies at $z_L =0.5$. Also we have in mind the fact  that condition $D_L \lll D_S$ should be met to use the approximation form of Equation \ref{equ:F_approx}.

\subsection{SNR and wave effects}\label{subsec:SNR}

%SNR plot for ET and discussion. How it looks like, where the SNR is greater than or close to 8. Are the wave effects visible?

With the estimated position and tension of the cosmic string, we can explore the wave effects further. In this section, we focus on the signal-to-noise ratio (SNR), $S/N$. We restrict ourselves to equal-mass binaries on circular orbits, placed at redshift $z_S=3$. This choice is motivated by the fact that the source should be much farther than the lens to justify the approximations used in Eq.~\ref{equ:F_approx}, and by the fact that most high-mass stellar black-hole mergers occur at high redshifts \citep{2013ApJ...779...72D}.
%The choice of mass span of single black hole, which vary from $m\in[5,50]$ solar mass, is justified by the fact that they lie in the detection range of the Einstein Telescope \citep{2022A&A...667A...2S}. 

For a mass range of $8$--$50\,M_\odot$, we simulate binary black-hole (BBH) mergers using \textsc{PyCBC} \citep{2024zndo..10473621N} and the \textsc{IMRPhenomD} waveform model \citep{2023ascl.soft07019P}. We then apply Eqs.~\ref{equ:F_approx} and \ref{equ:F_phased} to obtain the lensed waveform $\tilde{h}(f)$.

Given $\tilde{h}(f)$, we compute the SNR using \citep{2020MNRAS.495.2002L}
\begin{equation}
    \label{equ:SNR}
    \left(\frac{S}{N}\right)^2 = 4\int_{f_c^l}^{f_c^h}\frac{|\tilde{h}|^2}{S_h}\,df\
\end{equation}
Here, $f_c^l$ and $f_c^h$ are the lower and upper cut-off frequencies, respectively, and $S_h$ is the one-sided noise power spectral density of the detector. As the representative instrument, we choose the Einstein Telescope (ET-D) \citep{2010CQGra..27s4002P,2023JCAP...07..068B}. We assume that the signal is detected for $S/N\geq 8$ ($\log(S/N)\gtrsim 0.903$), a commonly used threshold \citep{Fairhurst:2010is,abbott_abbott_acernese_ackley_adams_adhikari_adhikari_adya_affeldt_agarwal_in._2023}.
%If this condition is fulfilled it is obvious that the investigated detector, in our case the Einstein Telescope, can detect it.
\begin{figure}
 \centering
    \begin{subfigure}{0.39\textwidth}
       \includegraphics[width=\textwidth]{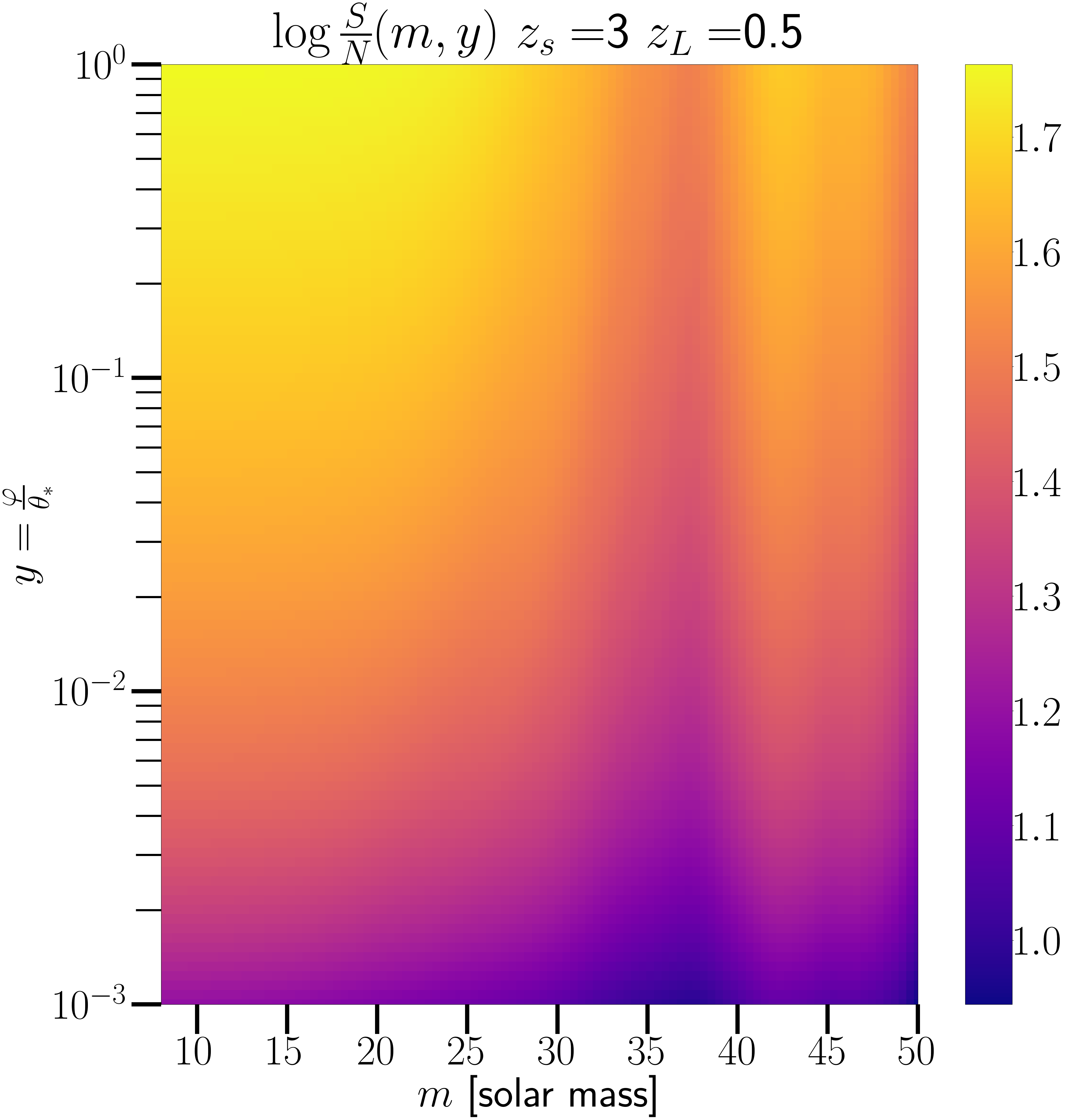}%
       \caption{$z_S=\mathrm{const}$}
       \label{fig:SNR_zs}
    \end{subfigure}
    
     \vspace{\floatsep}
    \begin{subfigure}{0.39\textwidth}
       \includegraphics[width=\textwidth]{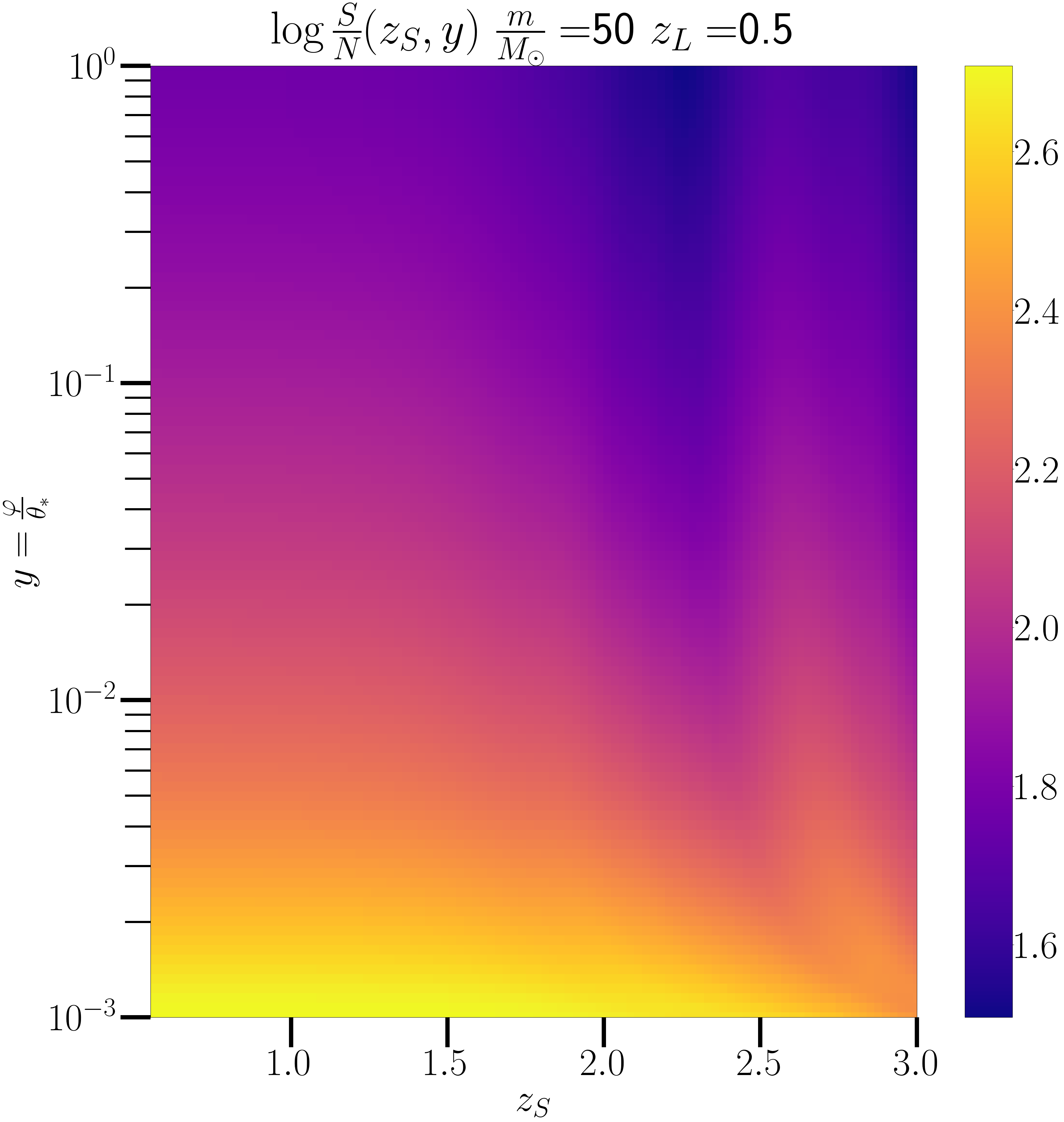}%
       \caption{$m=\mathrm{const}$}
       \label{fig:SNR_m}
    \end{subfigure}
    \caption{SNR heat maps. Panel (a) shows the single black-hole mass $m$ on the $x$-axis and the dimensionless source position $y$ on the $y$-axis; the colour bar indicates $\log(S/N)$. The calculation assumes $z_S=3$ and $z_L=0.5$. Panel (b) shows the same quantity as a function of $z_S$ and $y$ for a fixed mass $m=50\,M_\odot$.}
        \label{fig:SNR}
\end{figure}

We observe oscillations in Fig.~\ref{fig:SNR} that resemble the fringe structure in Fig.~\ref{fig:amp_2D}. These features are visible in Fig.~\ref{fig:SNR_m} for $z_S\gtrsim 2$ and in Fig.~\ref{fig:SNR_zs} for $m\gtrsim 30\,M_\odot$ and $y\gtrsim 0.01$.

In all panels, $S/N>8$, indicating that the Einstein Telescope is a promising instrument for detecting this type of event. The oscillatory behaviour in Fig.~\ref{fig:SNR} also reveals ``damped'' regions where the SNR is an order of magnitude lower, which may have consequences for searches based on multiple lensing events.

\subsection{Wave effects in waveforms and strains}\label{subsec:waveforms}

%In this subsection the plots of waveforms and strains will be presented. How the cosmic string lensing affects them for different masses and inclinations?

To illustrate how wave effects deform the waveforms, we present time-domain signals and characteristic strains.

We again use \textsc{PyCBC} to simulate unlensed waveforms, apply the same procedure as in Section~\ref{subsec:SNR} to obtain the lensed counterpart, and then transform the result to the time domain using the inverse Fourier transform. We also compare the characteristic strain of the simulated events with the strain sensitivities of aLIGO and Advanced Virgo \citep{2021MPLA...3630022C}, the Einstein Telescope, and Cosmic Explorer \citep{2019BAAS...51g..35R}.

\begin{figure*}
 \centering
    \begin{subfigure}{0.9\linewidth}
       \includegraphics[width=\textwidth]{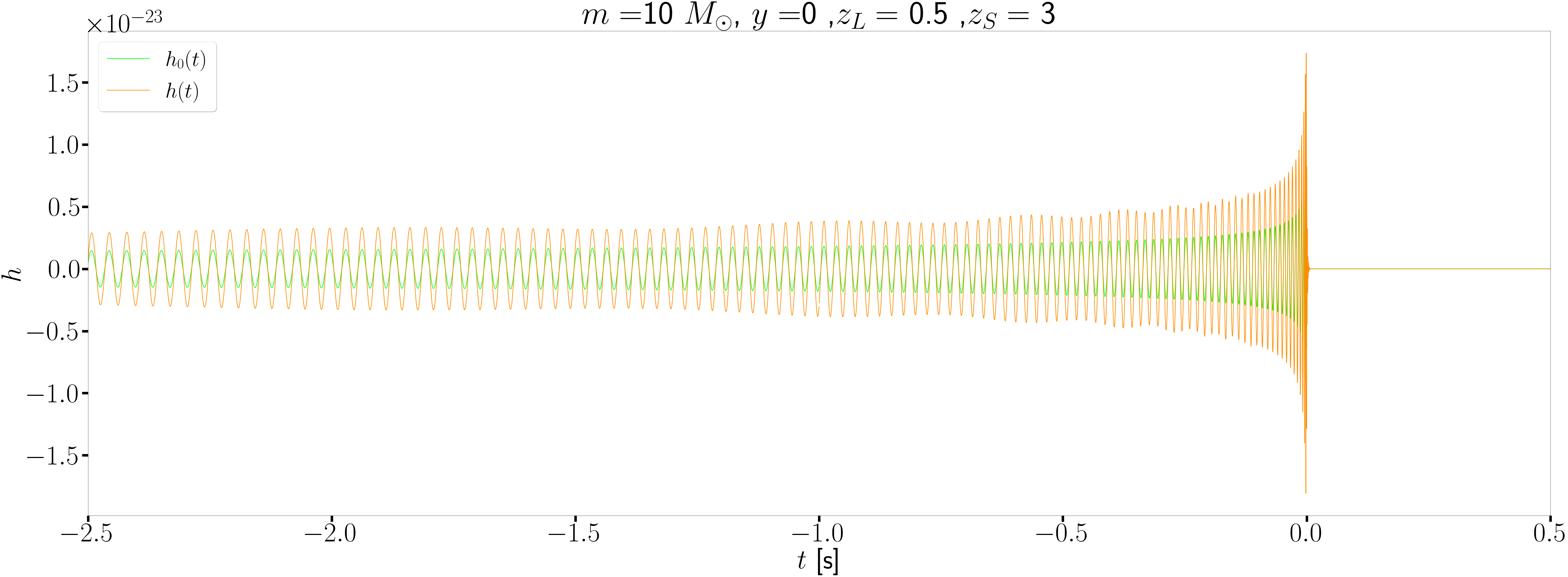}%
       \caption{$y=0$}
       \label{fig:hp_10_0}
    \end{subfigure}  
    
    \vspace{\floatsep}
    \begin{subfigure}{0.9\linewidth}
       \includegraphics[width=\textwidth]{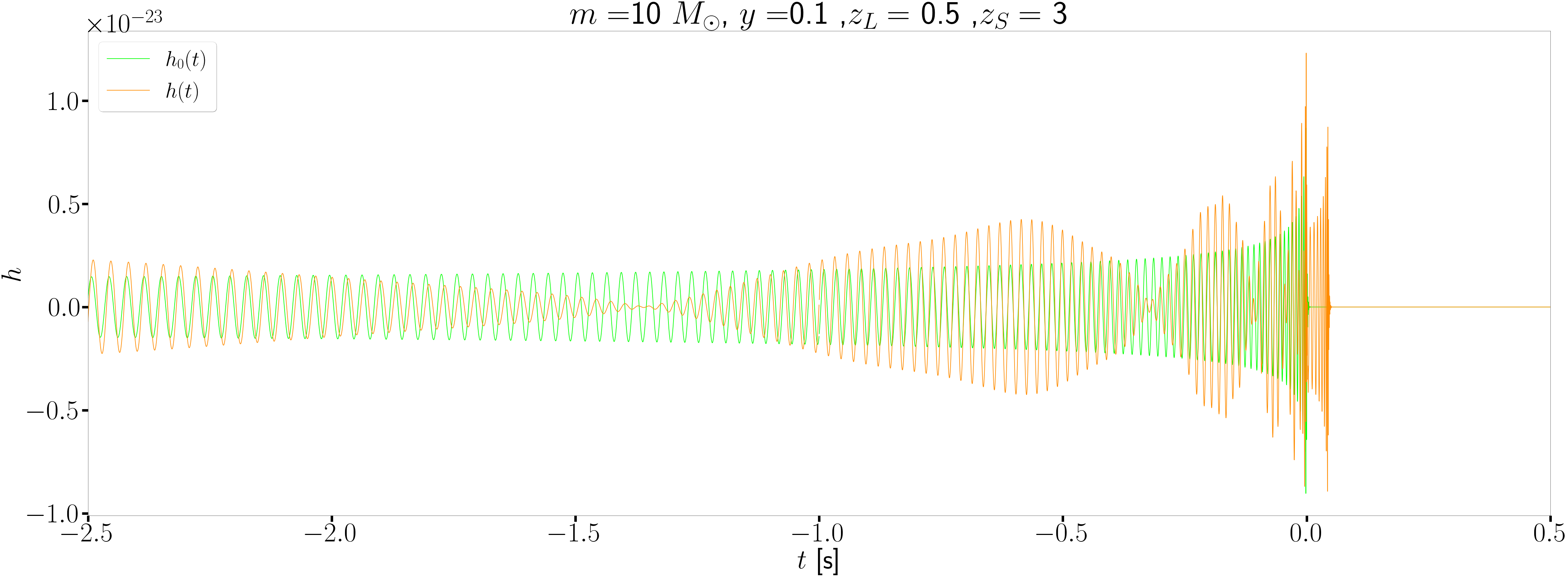}%
       \caption{$y=0.1$}
       \label{fig:hp_10_0.1}
    \end{subfigure}
    
      \vspace{\floatsep}
    \begin{subfigure}{0.9\linewidth}
       \includegraphics[width=\textwidth]{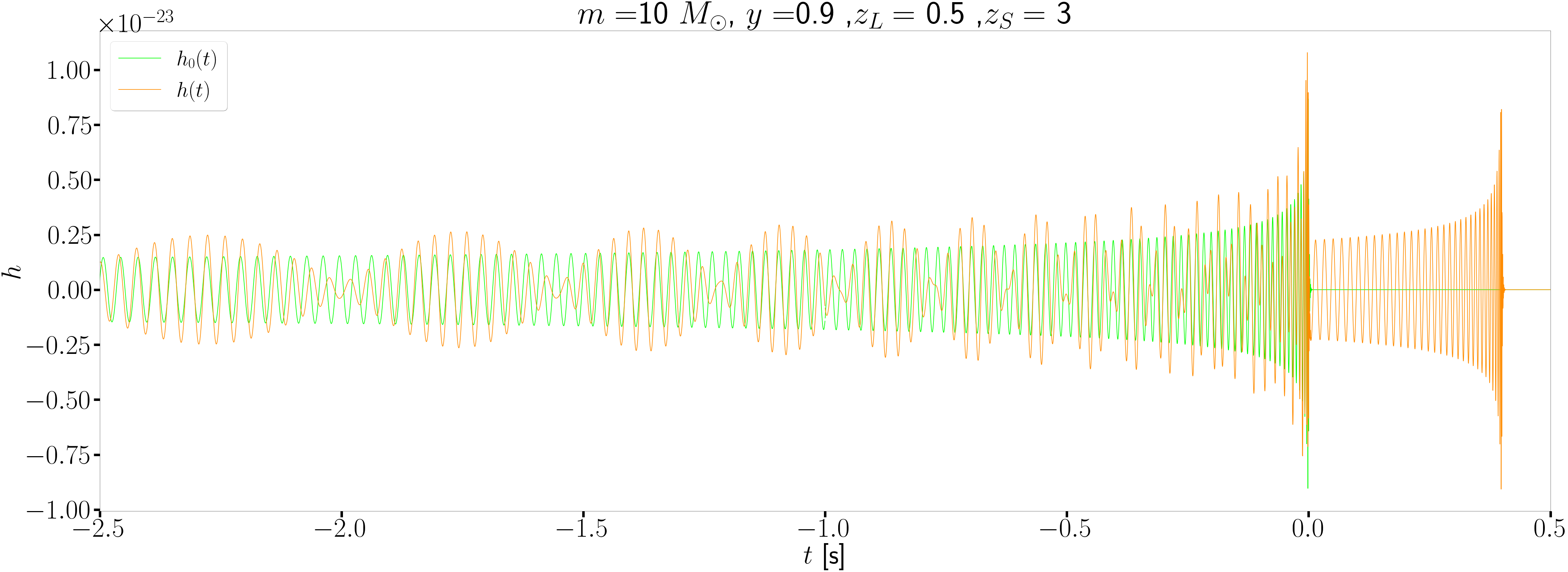}%
       \caption{$y=0.9$}
       \label{fig:hp_10_0.9}
    \end{subfigure}
    \caption{Time-domain waveforms for $m=10\,M_\odot$, with source redshift $z_S=3$ and lens redshift $z_L=0.5$. The $x$-axis shows time (s) and the $y$-axis shows strain $h(t)$. From top to bottom, the dimensionless source position is $y=0$, $0.1$, and $0.9$, respectively.}
    \label{fig:hp_10}

    \end{figure*}

As an example, we consider BBH mergers with component masses of $10\,M_\odot$. To show how lensing deforms the waveform, Fig.~\ref{fig:hp_10} compares the lensed waveform $h(t)$ with the unlensed one $h_0(t)$.

 \begin{figure}
 \centering
    \begin{subfigure}{0.39\textwidth}
       \includegraphics[width=\textwidth]{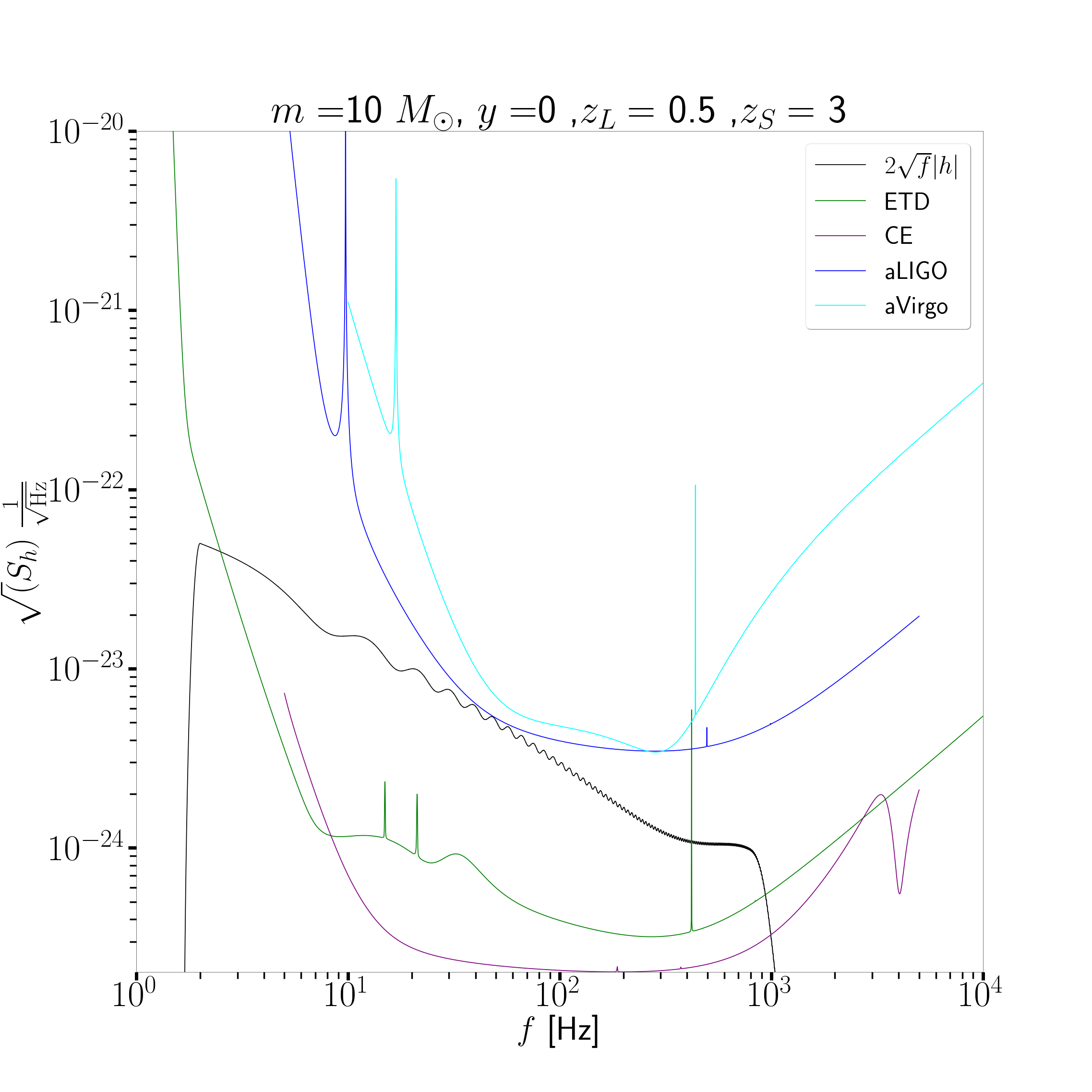}%
       \caption{$y=0$}
       \label{fig:strain_10_0}
    \end{subfigure}
    
      \vspace{\floatsep}
    \begin{subfigure}{0.39\textwidth}
       \includegraphics[width=\textwidth]{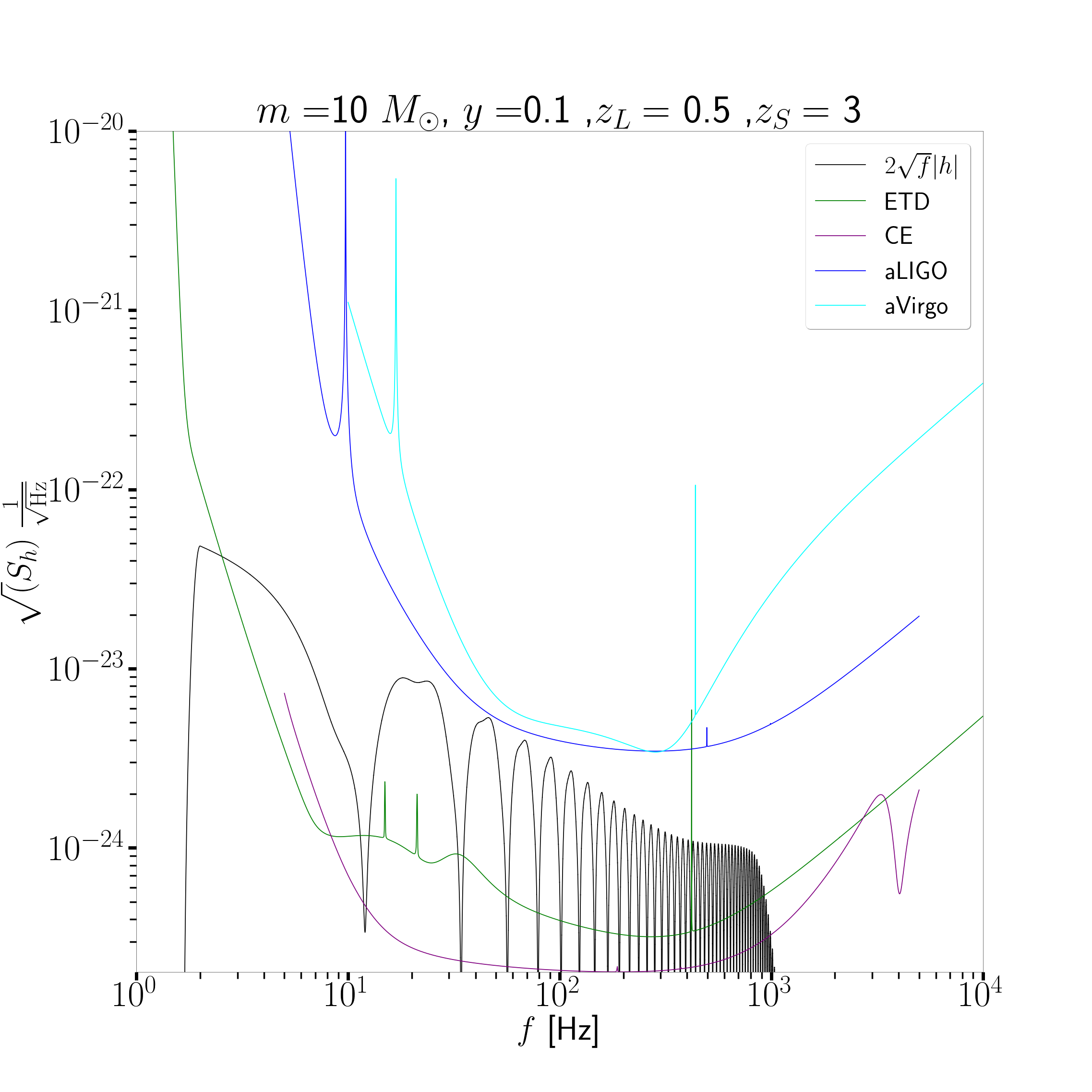}%
       \caption{$y=0.1$}
       \label{fig:strain_10_0.1}
    \end{subfigure}
    
      \vspace{\floatsep}
    \begin{subfigure}{0.39\textwidth}
       \includegraphics[width=\textwidth]{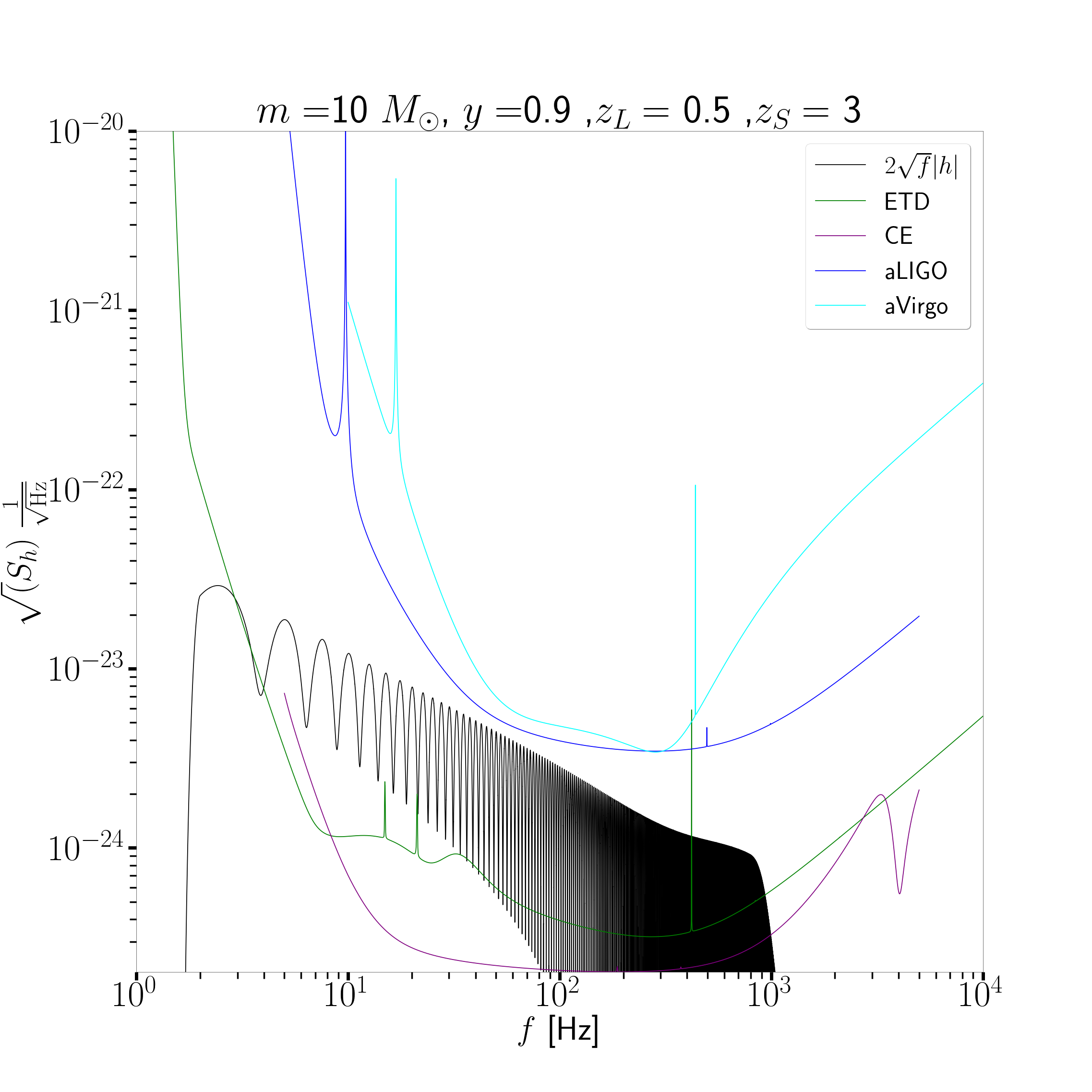}%
       \caption{$y=0.9$}
       \label{fig:strain_10_0.9}
    \end{subfigure}
    \caption{Characteristic strain for $m=10\,M_\odot$, with source redshift $z_S=3$ and lens redshift $z_L=0.5$. The black curve shows the strain of the merger; coloured curves show detector sensitivities (ET-D in green, Cosmic Explorer in violet, Advanced Virgo in cyan, and aLIGO in blue). The $x$-axis shows frequency $f$ (Hz) and the $y$-axis shows the characteristic strain. From top to bottom, the dimensionless source position is $y=0$, $0.1$, and $0.9$.}
     \label{fig:strain_10}

    \end{figure}

The top panel of Fig.~\ref{fig:hp_10_0} shows a lensing case with amplification only. The middle panel of Fig.~\ref{fig:hp_10_0.1} presents an intermediate case, where diffraction is clearly visible through characteristic beat patterns \citep{2021PhRvD.103d4005H}; weak interference is also present. Finally, the bottom panel of Fig.~\ref{fig:hp_10_0.9} shows a case with combined diffraction and strong interference, visible as the overlap of two distinct waveform images. This behaviour is characteristic of lensing. We also see that the separation between the wavefronts increases with $y$, consistent with Eq.~\ref{equ:time_delay}.

As before, to assess detectability, we compare the characteristic strain of the merger with the detector sensitivities. In Fig.~\ref{fig:strain_10}, the merger strain (black curve) barely reaches the sensitivity of aLIGO and Advanced Virgo, whereas it lies well within the sensitivity bands of the Einstein Telescope and Cosmic Explorer. This indicates that the event should be detectable with third-generation detectors. The beat patterns in the strain plots also become denser as $y$ increases.

The oscillatory behaviour seen in the beat patterns is also visible in the time-domain waveforms (Figs.~\ref{fig:hp_10_0.1} and \ref{fig:hp_10_0.9}) and in the last two strain panels (Figs.~\ref{fig:strain_10_0.1} and \ref{fig:strain_10_0.9}). It is present for $y=0$ as well (Fig.~\ref{fig:strain_10_0}), but it is barely visible. These features will be useful in the next section, where we estimate lens parameters using simulated events.

\section{Detectability of cosmic strings by their lensing with Einstein Telescope }\label{sec:ET_sim}

%New data of constraining cosmic string by multiple observations of lensing effects by Einstein Telescope. The data from Startrack simulation \citep{2020A&A...638A..94O} would be used. Next using \textsc{bilby} \citep{bilby_paper}  the constraints of lens could be estimated by MCMC \citep{bilby_mcmc_paper}. With the estimated parameters, we can build the distribution.

In this section, we discuss how the cosmic-string tension could be constrained using multiple mock observations of lensing with the Einstein Telescope. The section is divided into three subsections. First, we present the mock data. Next, we introduce the Bayesian inference technique. Finally, we present the results.

\subsection{Mock data}\label{subsec:mock}
To assess whether ET can measure the cosmic-string tension using multiple GW lensing detections, we require a realistic source population. We use the open-source catalogue from \citep{2020A&A...638A..94O}, based on the \textsc{StarTrack} stellar-evolution simulation \citep{2008ApJS..174..223B}. \textsc{StarTrack} is a rapid population-synthesis code: instead of solving the full stellar-structure equations for every star, it uses calibrated evolutionary prescriptions and fitting formulae to evolve large numbers of stellar binaries statistically. The code is widely used in GW astronomy to predict compact-object populations \citep{Perigois:2020ymr,2022A&A...667A...2S}. In \cite{Wiktorowicz_2019}, the \textsc{StarTrack} population was also examined in the context of optical lensing.

As in Subsection~\ref{subsec:model}, we place the string at $z_L=0.5$ with $\log(G\mu)=-10$. We randomly select eight BBH systems with source redshifts in the range $0.55\leq z_S\leq 4$. We also assign a fixed right ascension and a value of $y$ to each system, generated with a simple Monte Carlo routine written in \textsc{Python}.

Because of the \textit{mass-sheet degeneracy} \citep{2021PhRvD.104b3503C}, which for GW lensing by a cosmic string can make the tension degenerate with the lens distance, we first measure the time delay $\Delta t$ (Eq.~\ref{equ:time_delay}). From this, we can infer $y$ and $\mu_{\mathrm{eff}}$.

We calculate and use the following parameters:
\begin{itemize}
    \item $\mathcal{M}_c(1+z_S)$ --- redshifted chirp mass,
    \item $q$ --- mass ratio,
    \item $z_S$ --- source redshift,
    \item $\log(G\mu)$ --- logarithm of the string tension,
    \item $\log\chi_{\mathrm{eff}}$ --- logarithm of the effective comoving distance (Eq.~\ref{equ:chi_eff}),
    \item $\log\mu_{\mathrm{eff}}$ --- logarithm of the effective tension (Eq.~\ref{equ:mu_eff}),
    \item $y$ --- dimensionless source position (Eq.~\ref{equ:y}).
\end{itemize}

The chirp mass is given by \citep{2021A&A...649A..57O}
\begin{equation}
    \label{equ:chirp_mass}
    \mathcal{M}_c = \frac{(m_1m_2)^\frac{3}{5}}{(m_1+m_2)^\frac{1}{5}}\,
\end{equation}
Here, $m_i$ are the component masses and $z_S$ is the source redshift.
\begin{table*}[]
\centering
\begin{tabular}{|cccccccccc|}
\hline
\multicolumn{10}{|c|}{Exact parameters of population} \\ \hline
\multicolumn{1}{|c|}{ID} &
  \multicolumn{1}{c|}{$\mathcal{M}_c(1+z_S)$ [$M_\odot$]} &
  \multicolumn{1}{c|}{$q$} &
  \multicolumn{1}{c|}{$z_S$} &
  \multicolumn{1}{c|}{$\log{(G\mu)}$} &
  \multicolumn{1}{c|}{$\log{\chi_{\mathrm{eff}}}$ [$\log{(\mathrm{Mpc})}$]} &
  \multicolumn{1}{c|}{$\log{\mu_{\mathrm{eff}}}$ [$\log{(\mathrm{Mpc}^{\frac{1}{2}})}$]} &
  \multicolumn{1}{c|}{$\Delta t_{\mathrm{model}}$ [s]} &
  \multicolumn{1}{c|}{$y$} &
  $\Delta t_{\mathrm{obs}}$ [s] \\ \hline
\multicolumn{1}{|c|}{$3432056$} &
  \multicolumn{1}{c|}{$60.98$} &
  \multicolumn{1}{c|}{$0.75$} &
  \multicolumn{1}{c|}{$3.37$} &
  \multicolumn{1}{c|}{$-10.00$} &
  \multicolumn{1}{c|}{$3.14$} &
  \multicolumn{1}{c|}{$-8.43$} &
  \multicolumn{1}{c|}{$0.27$} &
  \multicolumn{1}{c|}{$0.59$} &
  $0.27$ \\ \hline
\multicolumn{1}{|c|}{$3178360$} &
  \multicolumn{1}{c|}{$119.24$} &
  \multicolumn{1}{c|}{$0.89$} &
  \multicolumn{1}{c|}{$2.35$} &
  \multicolumn{1}{c|}{$-10.00$} &
  \multicolumn{1}{c|}{$3.11$} &
  \multicolumn{1}{c|}{$-8.44$} &
  \multicolumn{1}{c|}{$0.41$} &
  \multicolumn{1}{c|}{$0.98$} &
  $0.41$ \\ \hline
\multicolumn{1}{|c|}{$2899582$} &
  \multicolumn{1}{c|}{$70.93$} &
  \multicolumn{1}{c|}{$0.77$} &
  \multicolumn{1}{c|}{$1.82$} &
  \multicolumn{1}{c|}{$-10.00$} &
  \multicolumn{1}{c|}{$3.08$} &
  \multicolumn{1}{c|}{$-8.46$} &
  \multicolumn{1}{c|}{$0.12$} &
  \multicolumn{1}{c|}{$0.31$} &
  $0.12$ \\ \hline
\multicolumn{1}{|c|}{$2863555$} &
  \multicolumn{1}{c|}{$87.47$} &
  \multicolumn{1}{c|}{$0.9$} &
  \multicolumn{1}{c|}{$1.77$} &
  \multicolumn{1}{c|}{$-10.00$} &
  \multicolumn{1}{c|}{$3.07$} &
  \multicolumn{1}{c|}{$-8.46$} &
  \multicolumn{1}{c|}{$0.25$} &
  \multicolumn{1}{c|}{$0.65$} &
  $0.25$ \\ \hline
\multicolumn{1}{|c|}{$2402527$} &
  \multicolumn{1}{c|}{$40.35$} &
  \multicolumn{1}{c|}{$0.79$} &
  \multicolumn{1}{c|}{$1.34$} &
  \multicolumn{1}{c|}{$-10.00$} &
  \multicolumn{1}{c|}{$3.02$} &
  \multicolumn{1}{c|}{$-8.49$} &
  \multicolumn{1}{c|}{$0.16$} &
  \multicolumn{1}{c|}{$0.46$} &
  $0.16$ \\ \hline
\multicolumn{1}{|c|}{$2302658$} &
  \multicolumn{1}{c|}{$52.99$} &
  \multicolumn{1}{c|}{$0.69$} &
  \multicolumn{1}{c|}{$1.28$} &
  \multicolumn{1}{c|}{$-10.00$} &
  \multicolumn{1}{c|}{$3.00$} &
  \multicolumn{1}{c|}{$-8.5$} &
  \multicolumn{1}{c|}{$0.13$} &
  \multicolumn{1}{c|}{$0.39$} &
  $0.13$ \\ \hline
\multicolumn{1}{|c|}{$2075086$} &
  \multicolumn{1}{c|}{$62.34$} &
  \multicolumn{1}{c|}{$0.98$} &
  \multicolumn{1}{c|}{$1.07$} &
  \multicolumn{1}{c|}{$-10.00$} &
  \multicolumn{1}{c|}{$2.95$} &
  \multicolumn{1}{c|}{$-8.53$} &
  \multicolumn{1}{c|}{$0.26$} &
  \multicolumn{1}{c|}{$0.92$} &
  $0.26$ \\ \hline
\multicolumn{1}{|c|}{$2056987$} &
  \multicolumn{1}{c|}{$46.53$} &
  \multicolumn{1}{c|}{$0.76$} &
  \multicolumn{1}{c|}{$1.04$} &
  \multicolumn{1}{c|}{$-10.00$} &
  \multicolumn{1}{c|}{$2.94$} &
  \multicolumn{1}{c|}{$-8.53$} &
  \multicolumn{1}{c|}{$0.13$} &
  \multicolumn{1}{c|}{$0.48$} &
  $0.13$ \\ \hline
\end{tabular}
\caption{Exact values of parameters from \textsc{StarTrack} simulation. In the first column we have an event's unique ID. Next in the second we have the redshifted chirp mass $\mathcal{M}_c(1+z_S)$  in solar masses. Next is mass ratio $q$. Fourth is  redshift of the source $z_S$. In the fifth, the sixth and the seventh column ,are logarithms of the string's tension $\log{(G\mu)}$, the effective distance $\log{\chi_{\mathrm{eff}}}$ and tension $\log{\mu_{\mathrm{eff}}}$. Next we have the exact time delay $\Delta t_{\mathrm{model}}$ in seconds. In the  ninth  column we have the dimensionless inclination $y$. The last column represents measured time delay  $\Delta t_{\mathrm{obs}}$ in seconds. }
\label{tab:exact}
\end{table*}

So in the rest of the work we will focus on fitting the following set of parameters:
\begin{itemize}
    \item{In the first part $\theta_1\in \{ \mathcal{M}_c(1+z_S),q,z_S,y,\log{\mu_{\mathrm{eff}}}\}$}
    \item{In the second part $\theta_2\in\{ \log{(G\mu)},\log{\chi_{\mathrm{eff}}}\}$}
    \item{In the third part based on the priors from all 8 previous estimations  $\theta_3\in\{ \log{(G\mu)}\}$}
\end{itemize}

\subsection{Method}\label{subsec:bayesian}

In this subsection we will briefly discuss the methods used in next parts of estimation procedure. 

First we will discuss the fundamentals of Bayesian Inference. This method, based on the Bayes theorem, is widely used in the GW astronomy. The whole introduction is based on work \cite{2019PASA...36...10T}. According to Bayes theorem, the posterior distribution is expressed as
\begin{equation}
\label{equ:bayes}
p(\theta|d)=\frac{\mathcal{L}(d|\theta)\pi(\theta)}{\mathcal{Z}}
\end{equation}
where: $d$ - data; $\theta$ - set of model parameters; $\mathcal{L}$— likelihood; $\pi$— posterior probability density; $\mathcal{Z}$- normalization factor, also called \textit{Bayesian evidence}, equal to

\begin{equation}
\label{equ:bayesian_evidence}
{\mathcal{Z}} = \int d\theta\mathcal{L}(d|\theta)\pi(\theta)
\end{equation}

Because of the complexity of our task we must divide a standard likelihood into two parts as follows:

\begin{equation}
\label{equ:likelihood_division}
\log\mathcal L=\log\mathcal L_{\mathrm{GW}}+\log\mathcal L_{\Delta t}
\end{equation}
where $\mathcal L_{\mathrm{GW}}$ - GW strain likelihood; $\mathcal L_{\Delta t}$ - Time-delay likelihood, which can be written as :

\begin{equation}
    \label{equ:likelihood_delta_t}
    \log\mathcal L_{\Delta t}=-\frac12\left(\frac{\Delta t_{\mathrm{model}}-\Delta t_{\mathrm{obs}}}{\sigma_{\Delta t}}\right)^2
\end{equation}
where we take standard deviation equal to $\sigma_{\Delta t}=0.05\Delta t_{\mathrm{obs}}$. 

Next we will discuss the \textit{Nested Sampling} . First described in \citep{10.1214/06-BA127}, the key concept of this algorithm is transforming the integral, from Equation \ref{equ:bayesian_evidence} , over the parameter space $\theta$ into an integral over the "likelihood volume." Next samples with the lowest likelihood are excluded, replacing them with new ones with higher ones, effectively mapping the distribution structure. In our work we use \textsc{UltraNest} \citep{2021JOSS....6.3001B} and \textsc{Dynesty}  \citep{sergey_koposov_2025_17268284} samplers.

Because we want to break the mass-sheet degeneracy, we need to have multiple measurements of $\log{(G\mu)}$. To extract it from $\log{\mu_{\mathrm{eff}}}$ we use method called \textit{hyperinference} \citep{2017MNRAS.471.2801S}. Its key concept is to draw the event parameters ($\theta$) themselves from a population distribution ($\lambda$), in this way $\theta_i \sim p(\theta|\lambda) $.

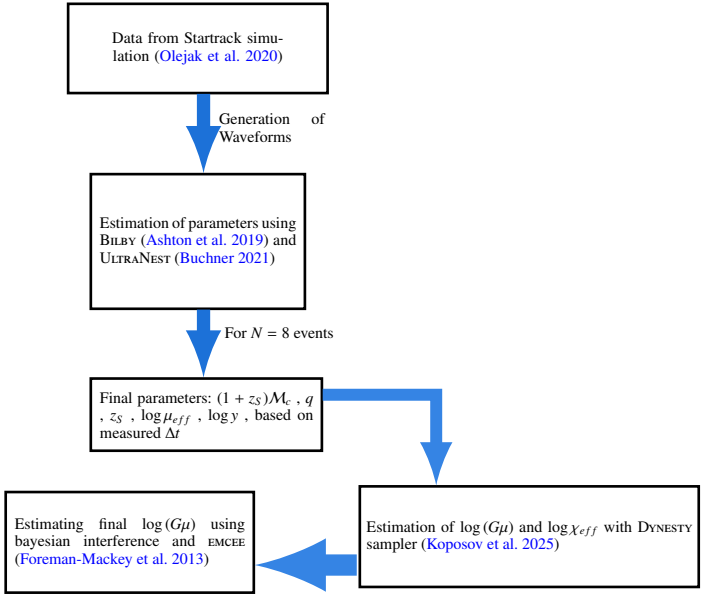
\begin{figure}
    \centering
\begin{tikzpicture}[transform shape, scale=0.6]
	% Paths, nodes and wires:
	\node[shape=rectangle, draw, line width=1pt, minimum width=5.715cm, minimum height=1.965cm] at (10.75, 8.125){} node[anchor=center, align=center, text width=5.327cm, inner sep=6pt] at (10.75, 8.125){Data from Startrack simulation \citep{2020A&A...638A..94O}};
	\path[draw={rgb,255:red,28;green,113;blue,216}, line width=5pt, -latex] (10.875, 7.125) -- (10.875, 5.375);
	\node[shape=rectangle, minimum width=2.715cm, minimum height=1.215cm] at (12.375, 6.375){} node[anchor=west, align=justify, text width=2.327cm, inner sep=6pt] at (11, 6.375){Generation of Waveforms};
	\node[shape=rectangle, draw, line width=1pt, minimum width=4.715cm, minimum height=2.965cm] at (10.75, 3.875){} node[anchor=west, align=justify, text width=4.327cm, inner sep=6pt] at (8.375, 3.875){Estimation of parameters using \textsc{Bilby} \citep{bilby_paper} and  \textsc{UltraNest} \citep{2021JOSS....6.3001B}};
	\path[draw={rgb,255:red,28;green,113;blue,216}, line width=5pt, -latex] (10.875, 2.375) -- (10.875, 0.875);
	\node[shape=rectangle, minimum width=3.465cm, minimum height=0.965cm] at (12.875, 1.875){} node[anchor=west, align=justify, text width=3.077cm, inner sep=6pt] at (11.125, 1.875){For $N=8$ events};
	\node[shape=rectangle, draw, line width=1pt, minimum width=5.09cm, minimum height=1.59cm] at (10.937, 0.062){} node[anchor=west, align=justify, text width=4.702cm, inner sep=6pt] at (8.375, 0.062){Final parameters: $(1+z_S)\mathcal{M}_c$ , $q$ , $z_S$ , $\log{\mu_{eff}}$ , $\log{y}$ , based on measured $\Delta t$};
	\node[shape=rectangle, draw, line width=1pt, minimum width=7.465cm, minimum height=2.215cm] at (18.05, -2.625){} node[anchor=west, align=justify, text width=7.177cm, inner sep=6pt] at (14.25, -2.625){Estimation of $\log{(G\mu})$ and $\log{\chi_{eff}}$ with \textsc{Dynesty} sampler \citep{sergey_koposov_2025_17268284}};
	\node[shape=rectangle, draw, line width=1pt, minimum width=5.465cm, minimum height=2.215cm] at (9.25, -2.75){} node[anchor=west, align=justify, text width=5.077cm, inner sep=6pt] at (6.5, -2.75){Estimating final $\log{(G\mu)}$ using bayesian interference and \textsc{emcee} \citep{2013PASP..125..306F}};
	\path[draw={rgb,255:red,53;green,132;blue,228}, line width=4.5pt, -latex] (13.5, 0.5) -- (16, 0.5) -- (16, -1.5);
	\path[draw={rgb,255:red,53;green,132;blue,228}, line width=8pt, -latex] (14.25, -3.25) -- (12, -3.25);
\end{tikzpicture}

\caption{Pipeline of code used to estimate the simulated events with \textsc{Bilby}}
    \label{fig:bilby_pipeline}

\end{figure}

We organize our estimation as follows:
\begin{enumerate}
    \item{Using Bilby \citep{bilby_paper} we generate an injections using \textsc{IRPhenomXPHM} waveform model \citep{2023ascl.soft07019P} for each data from Table \ref{tab:exact} }
    \item{Using UltraNest \citep{2021JOSS....6.3001B} we estimate set of parameters $\theta_1$. We take likelihood from Equation \ref{equ:likelihood_division};}
    \item{With Dynesty \citep{sergey_koposov_2025_17268284} we break $\log{\mu_{\mathrm{eff}}}$ based on Equation \ref{equ:mu_eff};}
    \item{At last using emcee we hyperinfere final string tension $\bar{\log{(G\mu)}}$.}
\end{enumerate}
The whole concept of whole procedure is summarized in graph in Figure \ref{fig:bilby_pipeline}.

\subsection{Estimation procedure and result}\label{subsec:result}
Parameters from Table \ref{tab:exact} are used as injections. With \textsc{bilby} \citep{bilby_paper, bilby_mcmc_paper}, Bayesian inference \textsc{Python} package for for gravitational wave astronomy, we can simulate detection of lensed merger with Gaussian noise and lower cut-off frequency $f_{\mathrm{low}}=5 $ Hz. To make the calculation easier we impose add $\Delta t$ to our likelihood 

In the first part  we use the following priors, which bounds are as follows:
\begin{itemize}
    \item{$\mathcal{M}_c(1+z_S)\in[10,150]$ $M_\odot$-uniform;}
    \item{$q\in[0.6,1]$-uniform;}
    \item{$z_S\in[0.25,4]$-uniform;}
    \item{$\log{\mu_{\mathrm{eff}}}\in[-11, -7]$-uniform;}
    \item{$y\in[0.3,1]$-loguniform;}
\end{itemize}
Rest we set fixed. We use \textsc{UltraNest} \citep{2021JOSS....6.3001B} as a sampler.

In the second part to estimate $\log{(G\mu)}$ and $\log{\chi_{\mathrm{eff}}}$ with the following priors:
\begin{itemize}
    %\item{$\mathcal{M}_c(1+z_S)\in[10,150]$ $M_\odot$-uniform;}
    %\item{$q\in[0.6,1]$-uniform;}
    %\item{$z_S\in[0.25,4]$-uniform;}
    \item{$\log{\chi_{\mathrm{eff}}}\in[2, \frac{\chi_S}{4}]$-uniform;}
    \item{$\log{(G\mu)}\in[-11, -9]$-uniform;}
\end{itemize}
At last as a hyperpopulation we use priors for each individual $\log{(G\mu)}$. To compute final parameters with errors in all cases we use $90\%$ credible interval.

\begin{figure}
    \centering
    \includegraphics[width=0.95\linewidth]{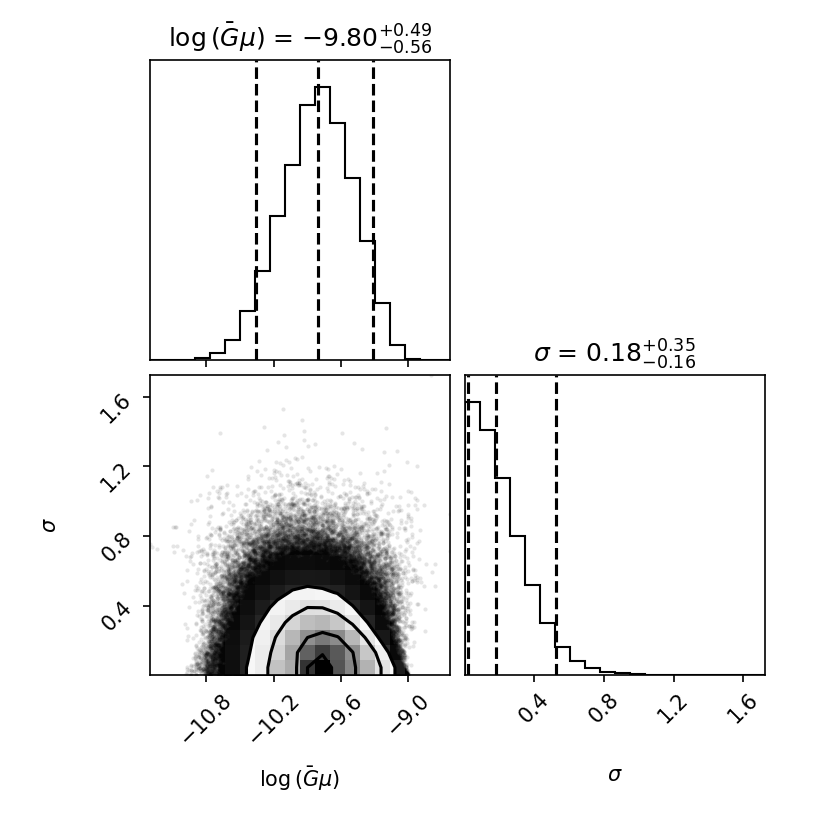}
    \caption{Corner plot of estimated mean cosmic string tension based on all eight events. The upper plot present distribution of $\bar{\log{(G\mu)}}$, which is based on estimated priors. As we can see its close to normal, and estimated final parameter $\bar{\log{(G\mu)}}=-9.80^{+0.49}_{-0.56} $ is consistent with injected value equal to $-10$. Lower the 2D distribution of previously mentioned parameter and $\sigma_{\log{(G\mu)}}$ - the standard deviation, which single distribution lies on the lower right. From it we can see that it is close to $0$, because all injections, had the same value of string tension.   }
    \label{fig:hyper_tension}
\end{figure}

After applying the whole pipeline, we get final cosmic string tension, which is presented in Figure \ref{fig:hyper_tension}. As we can see the $\bar{\log{(G\mu)}}$ is really close to real value equal $-10$ and lies within the errors. Additionally  dispersion is small, which implies that the individual estimations are consistent.

\begin{table*}[]
\centering
\begin{tabular}{|cccccc|}
\hline
\multicolumn{6}{|c|}{Recovered parameters of population} \\ \hline
\multicolumn{1}{|c|}{ID} &
  \multicolumn{1}{c|}{$\mathcal{M}_c(1+z_S)$ [$M_\odot$]} &
  \multicolumn{1}{c|}{$q$} &
  \multicolumn{1}{c|}{$z_S$} &
  \multicolumn{1}{c|}{$\log{\mu_{\mathrm{eff}}}$ [$\log{(\mathrm{Mpc}^{\frac{1}{2}})}$]} &
  $y$ \\ \hline
\multicolumn{1}{|c|}{$3432056$} &
  \multicolumn{1}{c|}{$61.044^{+0.063}_{-0.063}$} &
  \multicolumn{1}{c|}{$0.783^{+0.05}_{-0.038}$} &
  \multicolumn{1}{c|}{$3.38^{+0.11}_{-0.10}$} &
  \multicolumn{1}{c|}{$-8.4302^{+0.0025}_{-0.0025}$} &
  $0.5947^{+0.01}_{-0.019}$ \\ \hline
\multicolumn{1}{|c|}{$3178360$} &
  \multicolumn{1}{c|}{$119.28^{+0.11}_{-0.26}$} &
  \multicolumn{1}{c|}{$0.913^{+0.077}_{-0.094}$} &
  \multicolumn{1}{c|}{$2.324^{+0.045}_{-0.043}$} &
  \multicolumn{1}{c|}{$-8.4464^{+0.0024}_{-0.0024}$} &
  $0.989^{+0.011}_{-0.011}$ \\ \hline
\multicolumn{1}{|c|}{$2899582$} &
  \multicolumn{1}{c|}{$70.881^{+0.051}_{-0.047}$} &
  \multicolumn{1}{c|}{$0.744^{+0.02}_{-0.017}$} &
  \multicolumn{1}{c|}{$1.852^{+0.028}_{-0.028}$} &
  \multicolumn{1}{c|}{$-8.4616^{+0.0017}_{-0.0015}$} &
  $0.3139^{+0.0022}_{-0.0025}$ \\ \hline
\multicolumn{1}{|c|}{$2863555$} &
  \multicolumn{1}{c|}{$87.494^{+0.028}_{-0.065}$} &
  \multicolumn{1}{c|}{$0.932^{+0.059}_{-0.058}$} &
  \multicolumn{1}{c|}{$1.786^{+0.021}_{-0.021}$} &
  \multicolumn{1}{c|}{$-8.4634^{+0.0021}_{-0.0002}$} &
  $0.6456^{+0.0006}_{-0.0064}$ \\ \hline
\multicolumn{1}{|c|}{$2402527$} &
  \multicolumn{1}{c|}{$40.3456^{+0.0084}_{-0.0075}$} &
  \multicolumn{1}{c|}{$0.783^{+0.014}_{-0.012}$} &
  \multicolumn{1}{c|}{$1.363^{+0.017}_{-0.018}$} &
  \multicolumn{1}{c|}{$-8.4915^{+0.0003}_{-0.0024}$} &
  $0.4627^{+0.0052}_{-0.0006}$ \\ \hline
\multicolumn{1}{|c|}{$2302658$} &
  \multicolumn{1}{c|}{$53.005^{+0.016}_{-0.015}$} &
  \multicolumn{1}{c|}{$0.6994^{+0.0085}_{-0.0084}$} &
  \multicolumn{1}{c|}{$1.27^{+0.013}_{-0.014}$} &
  \multicolumn{1}{c|}{$-8.4982^{+0.0004}_{-0.0004}$} &
  $0.3857^{+0.0007}_{-0.0008}$ \\ \hline
\multicolumn{1}{|c|}{$2075086$} &
  \multicolumn{1}{c|}{$62.331^{+0.01}_{-0.02}$} &
  \multicolumn{1}{c|}{$0.941^{+0.053}_{-0.044}$} &
  \multicolumn{1}{c|}{$1.063^{+0.011}_{-0.01}$} &
  \multicolumn{1}{c|}{$-8.5267^{+0.0007}_{-0.0006}$} &
  $0.9162^{+0.0003}_{-0.0004}$ \\ \hline
\multicolumn{1}{|c|}{$2056987$} &
  \multicolumn{1}{c|}{$46.531^{+0.0091}_{-0.0092}$} &
  \multicolumn{1}{c|}{$0.7555^{+0.0089}_{-0.0089}$} &
  \multicolumn{1}{c|}{$1.0421^{+0.0097}_{-0.0097}$} &
  \multicolumn{1}{c|}{$-8.5334^{+0.0024}_{-0.0002}$} &
  $0.4813^{+0.0006}_{-0.0052}$ \\ \hline
\end{tabular}
\caption{Estimated values of parameters from the first step . In the first column we have event's unique ID. Next in second we have redshifted chirp mass $\mathcal{M}_c(1+z_S)$  in solar masses. Next is mass ratio $q$. Fourth is source redshift $z_S$. In fifth we have the logarithm of string's effective tension $\log{\mu_{\mathrm{eff}}}$. In the last  column we have a logarithm of the dimensionless inclination $y$.}
\label{tab:results_1}
\end{table*}
Parameters computed in first stage of estimation are placed in Table \ref{tab:results_1}. As we can see all of them are in good agreement with initial data from Table \ref{tab:exact}, which tells us that in case of lensing by cosmic string the merger parameters can be estimated correctly.
\section{Conclusions}\label{sec:disc_summ}

%The end of article. The discussion of obtained results. 

In this article we explored the possibilities of Einstein Telescope to detect cosmic strings and estimate its parameters.  In first part of the paper, we presented theoretical framework, based on precious works about this phenomena. In the second the wave effects were studied with plots for fixed positions of sources and lens. We can see that signals are efficiently strong to be detected by examined detector from Fig \ref{fig:SNR} . The wave effects are visible in waveforms (Fig. \ref{fig:hp_10}) and strains (Fig. \ref{fig:strain_10}).  In the third part the population from \textsc{StarTrack} simulation was used, to construct mock lensing events. With them we could fit parameters with nested sampling for individual phenomena. At last using hyperinference we have estimated string tension  with posterior from all events. The final result were equal to $\bar{\log{G\mu}}=-9.80^{+0.49}_{-0.56} $, which were consistent with injected value equal to $-10$. 

Based on our discussion done in previous paragraphs, we can conclude that Einstein Telescope is an excellent tool for detection and examination of cosmic string lensing events, even for such low tension values. Also in the future it will be beneficial to combine researched events with sources other than black hole binary mergers, such as neutron stars, Gravitational Wave Background radiation or electromagnetic ones. This could provide new constraints for examining such types of events. Also the possibilities of a Machine Learning techniques should be checked in this case. 
\FloatBarrier

\bibliography{sample}
%%%%%%%%%%%%%%%%%%%%%%%%%%%%%%%%%%%%%%%%%%%%%%%%%%%%%%%%%%%%%%

%%%%%%%%%%%%%%%%%%%%%%%%%%%%%%%%%%%%%%%%%%%%%%%%%%%%%%%%%%%%%%%
% Appendices must be placed after   \end{thebibliography}
% They will be placed automatically on a new page.
%%%%%%%%%%%%%%%%%%%%%%%%%%%%%%%%%%%%%%%%%%%%%%%%%%%%%%%%%%%%%%%

\end{document}